\documentclass[11pt,twoside]{article}
\usepackage{asp2010}

\resetcounters

\begin{document}

\title{AstroCloud, a Cyber-Infrastructure for Astronomy Research: Data Archiving and Quality Control}
\author{Boliang He$^1$, Chenzhou Cui$^1$, Dongwei Fan$^1$, Changhua Li$^1$, Jian Xiao$^2$, Ce Yu$^2$, Chuanjun Wang$^3$, Zihuang Cao$^1$, Junyi Chen$^2$, Weimin Yi$^3$, Shanshan Li$^1$, Linying Mi$^1$ and Sisi Yang$^1$}
\affil{$^1$National Astronomical Observatories, Chinese Academy of Sciences (CAS), 20A Datun Road, Beijing 100012, China\\
$^2$Tianjin University, 92 Weijin Road, Tianjin 300072, China\\
$^3$Yunnan Astronomical Observatory, CAS, P.0.Box110, Kunming 650011, China
}

\begin{abstract}
AstroCloud is a cyber-Infrastructure for Astronomy Research initiated by Chinese Virtual Observatory (China-VO) under funding support from NDRC (National Development and Reform commission) and CAS (Chinese Academy of Sciences)\footnote{\url{http://astrocloud.china-vo.org}}\citep{O8-5_Cui_adassxxiv}. To archive the astronomical data in China, we present the implementation of the astronomical data archiving system (ADAS). Data archiving and quality control are the infrastructure for the AstroCloud. Throughout the data of the entire life cycle, data archiving system standardized data, transferring data, logging observational data, archiving ambient data, And storing these data and metadata in database. Quality control covers the whole process and all aspects of data archiving.
\end{abstract}

\section{Introduction}

There are tens of telescopes running in China. Every night and day, they are producing several terabytes data. To archive these huge data and manage them, we present an implementation of an Astronomical Data Archiving System (ADAS). The data types which would be archived are the observation data and ambient data. The observation data such as image FITS, spectra FITS and observation log, are produced by telescope and data reduce pipeline. Ambient data are some environment data, such as weather, seeing data and allsky camera images.

Archived data is stored into the observatory¡¯ data center first, then Data transferred to AstroCloud data center via ADAS. In AstroCloud, we build a Data Access API For users and programs to access data. The following telescopes have been already using this archiving system to archive their data. These telescope are located in multiple sites in China: Guo Shoujing Telescope (LAMOST), Lijiang GMG 2.4m Telescope, Xinglong 2.16m Telescope, Delingha 50Bin Telescope, Huairou Solar Radio Telescope, Huairou Solar Multi-Channel Telescope and Fuxian 1m New Vacuum Solar Telescope (NVST).

\begin{figure}
  \centering
  \includegraphics[width=0.70\textwidth]{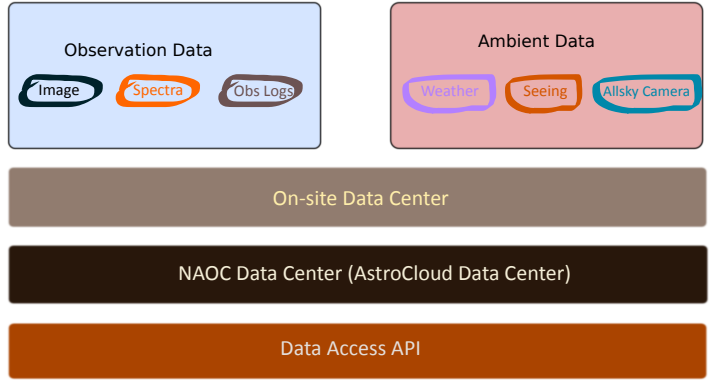}\\
  \caption{Data archive Framework}\label{fig1}
\end{figure}

\section{Data Model}

The type of ¡°raw¡± data include files and tables. FITS file mainly contain the raw data. FITS can be image, can be spectral, etc. The tables are catalog tables, ambient data tables, observational logs, etc.

Metadata consists of two types:

\begin{itemize}
  \item \textbf{Schema Metadata}: Schema Metadata stores all the databases, schemas, tables and columns information. The database-schema is similar to the IVOA TAP schemas\citep{IVOA_TAP}. 
  \item \textbf{Archive Metadata}: Archive Metadata stores the FITS files¡¯ header information. The must filed in database-schema is shown in Table \ref{tbl_amds}. Usually, One telescope has one table in archive database. 
\end{itemize}

\begin{table}[!ht]
	\caption{Archive Metadata database-schema}\label{tbl_amds}
	\smallskip
	\begin{center}
	{\small
	\begin{tabular}{lll}
		\tableline
		\noalign{\smallskip}
		\textbf{Column Name} & \textbf{Definition} & \textbf{Description} \\
		\noalign{\smallskip}
		\tableline
		\noalign{\smallskip}
		\texttt{id} 	& \texttt{SERIAL} & Auto increasing integer, Primary Key \\
		\texttt{filename} &	 \texttt{VARCHAR(30)} &	FITS file name \\
		\texttt{object} &	\texttt{VARCHAR(30)} &	Observation object \\
		\texttt{RA}	& \texttt{NUMERIC(12,8)} &	Right ascension, default J2000 \\
		\texttt{Dec} &	\texttt{NUMERIC(12,8)} &	Declination, default J2000 \\
		\texttt{filesize} &	\texttt{INTEGER} &	File size (bytes) \\
		\texttt{checksum} &	\texttt{VARCHAR(64)} &	MD5 checksum \\
		\texttt{recTime} &	\texttt{TIMESTAMP WITHOUT TIME ZONE} &	Recorded time \\
		\noalign{\smallskip}
		\tableline
	\end{tabular}
	}
	\end{center}
\end{table}

\section{Software archiving Architecture}

\begin{figure}
  \centering
  \includegraphics[width=0.70\textwidth]{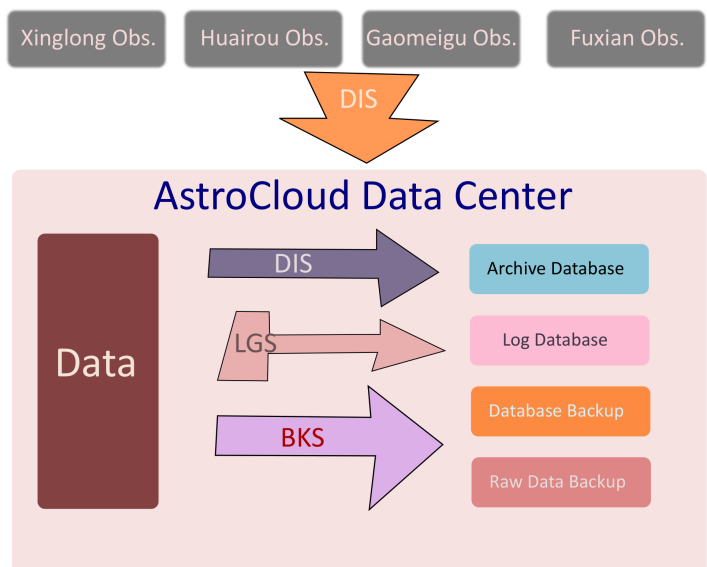}\\
  \caption{Software Architecture}\label{fig2}
\end{figure}

The system consists of four submodules \citep{2014PASP..126..674L}:

\begin{itemize}
  \item[1] \textbf{Data Transfer System (DTS)}. Data transferring is via network. The network transfer is scheduled. In the central data center in NAOC, we set up a Transfer Server to accept data transfer. We choose \texttt{rsync} tools running this service. Because it is open source and has a very good performances.  \citep{2009ASPC..411..540Z}
  
  \item[2] \textbf{Data Ingest System (DIS)}. DIS provides the data to database function. This procedure will parse the FITS header and choose the necessary filed to record into the database. We use the \texttt{AstroPy}\citep{AstroPy} to manipulate the FITS file, which can collect the FITS file header easily.\citep{2012SPIE.8451E..19D}

  \item[3] \textbf{Logging System (LGS)}. All the operation will be logged into the database. LGS is the procedure to log the operation: data transfer, data ingest, database replication, etc.
  
  \item[4] \textbf{Archive Backup System (BKS)}. BKS consists of files backup, database replication, and database backup. These operations are scheduled. 
\end{itemize}

\section{Archiving Pipeline}

\begin{itemize}
  \item[1] Data will be transferred to the data center in NAOC by DTS in schedule. 
  \item[2] After the data is finished transferred. DIS will start running, DIS will check the files¡¯ checksum, collect the FITS files¡¯ header and insert it into the archive database.
  \item[3] All the files has been checked and record into the database, gather these information (file amount, transfer log, database log, etc) to email these information to the system administrator and telescope operator.
  \item[4] These FITS files and database will be backup by BKS in schedule.
  \item[5] Database replication: archive database is the write-only database, the \texttt{SkyTools}\citep{SkyTools} replication procedure will replicate the database to the Query Databases for other user or system to access, such as Data Publish System\footnote{\url{http://explore.china-vo.org}}\citep{P1-3_Fan_adassxxiv}. 
\end{itemize}

\section{Quality Control}

Data quality can be controlled by the data archiving process. In DTS, every file has been made a MD5 checksum, before transferred and after transferred, transfer procedure will valid the checksum. Database is been checked and valid by schedule.

\section{Conclusions}

We developed and implemented an astronomical data archiving system that can be operated automatic.  When the data is produced, the procedure will be running quietly. When the procedure is finished, the operator will receive the job detail email.

\acknowledgements This paper is funded by National Natural Science Foundation of China (U1231108), Ministry of Science and Technology of China (2012FY120500), Chinese Academy of Sciences (XXH12503-05-05). Data resources are supported by Chinese Astronomical Data Center.

\bibliography{P5-1}

\begin{thebibliography}{}
\expandafter\ifx\csname natexlab\endcsname\relax\def\natexlab#1{#1}\fi
\expandafter\ifx\csname url\endcsname\relax
  \def\url#1{\texttt{#1}}\fi
\expandafter\ifx\csname urlprefix\endcsname\relax\def\urlprefix{URL }\fi
\providecommand{\eprint}[2][]{\url{#2}}

\bibitem[{IVO(2010)}]{IVOA_TAP}
 2010, {Table Access Protocol (TAP) Version 1.0}.
  \url{http://www.ivoa.net/documents/TAP/20100327/REC-TAP-1.0.html}

\bibitem[{Ast(2014)}]{AstroPy}
 2014, {AstroPy}. \url{http:///www.astropy.org/}

\bibitem[{Sky(2014)}]{SkyTools}
 2014, {SkyTools}. \url{http://pgfoundry.org/projects/skytools}

\bibitem[{{Cui} et~al.(2014){Cui}, {Yu}, {Xiao}, {He}, {Li}, {Fan}, {Wang},
  {Hong}, {Li}, {Mi}, {Wan}, {Cao}, {Wang}, {Yin}, {Fan}, {Wang}, \&
  {Yang}}]{O8-5_Cui_adassxxiv}
{Cui}, C., {Yu}, C., {Xiao}, J., {He}, B., {Li}, C., {Fan}, D., {Wang}, C.,
  {Hong}, Z., {Li}, S., {Mi}, L., {Wan}, W., {Cao}, Z., {Wang}, J., {Yin}, S.,
  {Fan}, Y., {Wang}, J., \& {Yang}, S. 2014, in ADASS XXIV, edited by A.~R.
  Taylor, \& J.~M. Stil (San Francisco: ASP), vol. TBD of ASP Conf. Ser., TBD

\bibitem[{{Dobrzycki} et~al.(2012){Dobrzycki}, {da Rocha}, {Vera}, {Vuong},
  {Bierwirth}, {Forch{\i}}, {Fourniol}, {Moins}, \&
  {Zampieri}}]{2012SPIE.8451E..19D}
{Dobrzycki}, A., {da Rocha}, C., {Vera}, I., {Vuong}, M.-H., {Bierwirth}, T.,
  {Forch{\i}}, V., {Fourniol}, N., {Moins}, C., \& {Zampieri}, S. 2012, in
  Society of Photo-Optical Instrumentation Engineers (SPIE) Conference Series,
  vol. 8451 of Society of Photo-Optical Instrumentation Engineers (SPIE)
  Conference Series

\bibitem[{{Fan} et~al.(2014){Fan}, {He}, {Xiao}, {Li}, {Li}, {Cui}, {Yu},
  {Hong}, {Yin}, {Wang}, {Cao}, {Fan}, {Mi}, {Wan}, \&
  {Wang}}]{P1-3_Fan_adassxxiv}
{Fan}, D., {He}, B., {Xiao}, J., {Li}, S., {Li}, C., {Cui}, C., {Yu}, C.,
  {Hong}, Z., {Yin}, S., {Wang}, C., {Cao}, Z., {Fan}, Y., {Mi}, L., {Wan}, W.,
  \& {Wang}, J. 2014, in ADASS XXIV, edited by A.~R. Taylor, \& J.~M. Stil (San
  Francisco: ASP), vol. TBD of ASP Conf. Ser., TBD

\bibitem[{{Laher} et~al.(2014){Laher}, {Surace}, {Grillmair}, {Ofek},
  {Levitan}, {Sesar}, {van Eyken}, {Law}, {Helou}, {Hamam}, {Masci},
  {Mattingly}, {Jackson}, {Hacopeans}, {Mi}, {Groom}, {Teplitz}, {Desai},
  {Hale}, {Smith}, {Walters}, {Quimby}, {Kasliwal}, {Horesh}, {Bellm},
  {Barlow}, {Waszczak}, {Prince}, \& {Kulkarni}}]{2014PASP..126..674L}
{Laher}, R.~R., {Surace}, J., {Grillmair}, C.~J., {Ofek}, E.~O., {Levitan}, D.,
  {Sesar}, B., {van Eyken}, J.~C., {Law}, N.~M., {Helou}, G., {Hamam}, N.,
  {Masci}, F.~J., {Mattingly}, S., {Jackson}, E., {Hacopeans}, E., {Mi}, W.,
  {Groom}, S., {Teplitz}, H., {Desai}, V., {Hale}, D., {Smith}, R., {Walters},
  R., {Quimby}, R., {Kasliwal}, M., {Horesh}, A., {Bellm}, E., {Barlow}, T.,
  {Waszczak}, A., {Prince}, T.~A., \& {Kulkarni}, S.~R. 2014, \pasp, 126, 674.
  \eprint{1404.1953}

\bibitem[{{Zampieri} et~al.(2009){Zampieri}, {Forchi}, {Gebbinck}, {Moins}, \&
  {Padovan}}]{2009ASPC..411..540Z}
{Zampieri}, S., {Forchi}, V., {Gebbinck}, M.~K., {Moins}, C., \& {Padovan}, M.
  2009, in Astronomical Data Analysis Software and Systems XVIII, edited by
  D.~A. {Bohlender}, D.~{Durand}, \& P.~{Dowler}, vol. 411 of Astronomical
  Society of the Pacific Conference Series, 540

\end{thebibliography}

\end{document}